\documentclass[11pt]{article}

\usepackage[english,francais]{babel}

\newfont{\ensmathquatorze}{msbm10 scaled 1400}
\newfont{\ensmathonze}{msbm10 scaled 1100}
\newfont{\ensmathdix}{msbm10}
\newfont{\ensmathneuf}{msbm10 scaled 833}
\newfont{\ensmathhuit}{msbm10 scaled 694}
\newfam\ensmathfam                        
\textfont\ensmathfam=\ensmathonze        
\scriptfont\ensmathfam=\ensmathdix       
\scriptscriptfont\ensmathfam=\ensmathhuit

\def\be{\begin{equation}}
\def\ee{\end{equation}}
\def\bea{\begin{eqnarray}}
\def\eea{\end{eqnarray}}

\newcommand\lab{l_{\alpha\beta}}

\newcommand\xab{x_{\alpha\beta}}
\newcommand\xba{x_{\beta\alpha}}

\newcommand\Gab{G_{(\alpha\beta )}}
\newcommand\Gba{G_{(\beta\alpha )}}
\newcommand\Gaa{G_{(ab )}}
\newcommand\Gbb{G_{(ba )}}
\newcommand\Gpaa{G'_{(ab )}}

\newcommand\Gpab{G'_{(\alpha\beta )}}
\newcommand\Gpba{G'_{(\beta\alpha )}}

\newcommand\dg{\partial_{\gamma}}
\newcommand\pal{\psi_{\alpha\beta}}
\newcommand\pbe{\psi_{\beta\alpha}}
\newcommand\ppal{\psi'_{\alpha\beta}}
\newcommand\ppbe{\psi'_{\beta\alpha}}
\newcommand\fal{\phi_{(\alpha\beta )}}
\newcommand\fbe{\phi_{(\beta\alpha )}}
\newcommand\fpal{\phi'_{(\alpha\beta )}}
\newcommand\fpbe{\phi'_{(\beta\alpha )}}

\newcommand\cab{\chi_{\alpha\beta}}
\newcommand\cba{\chi_{\beta\alpha}}
\newcommand\cpab{\chi'_{\alpha\beta}}
\newcommand\cpba{\chi'_{\beta\alpha}}

\newcommand\caa{\chi_{ab}}
\newcommand\cbb{\chi_{ba}}

\newcommand\cabi{\chi_{\alpha\beta_i}}

\newcommand\cpabi{\chi'_{\alpha\beta_i}}

\newcommand\sg{\sqrt {\gamma }}
\newcommand\pab{\psi_{ab}}
\newcommand\pba{\psi_{ba}}
\newcommand\ppab{\psi'_{ab}}
\newcommand\ppba{\psi'_{ba}}

\newcommand\dpl{ \frac{ \d \psi_{\alpha\beta}}{  \d x_{\alpha\beta}} }
\newcommand\dpt{ \frac{ \d \psi_{\beta\alpha}}{ \d x_{\alpha\beta}} }

\newcommand\wal{W_{\alpha\beta}}
\newcommand\wla{W_{\beta\alpha}}
\newcommand\wab{W_{ab}}

\newcommand\Wal{{\cal W}_{\alpha\beta}}
\newcommand\Wab{{\cal W}_{ab}}
\newcommand\Wla{{\cal W}_{\beta\alpha}}
\newcommand\Wba{{\cal W}_{ba}}
\def\d{{\rm d}}

\begin{document}

\selectlanguage{english}

\title{Spectral determinant on graphs with generalized boundary conditions}

\author{        Jean Desbois  }

\maketitle	

{\small
\noindent
Laboratoire de Physique Th\'eorique et Mod\`eles Statistiques.
Universit\'e Paris-Sud, B\^at. 100, F-91405 Orsay Cedex, France.

}

\begin{abstract}

 The spectral determinant of the Schr\"odinger operator
 ($ - \Delta + V(x) $) on a graph is
  computed for general boundary conditions. ($\Delta$  is the
 Laplacian and $V(x)$ is some potential defined on the graph).
  Applications to restricted random walks on graphs are discussed.

 \end{abstract}

\vskip1cm

\section{Introduction}\label{s1}

  The study of 
 spectral properties of the Laplacian operator on finite graphs began about
 fifty years ago. Many different domains are interested in the knowledge 
  of those properties - let us simply mention organic molecules \cite{ru},
    superconducting networks
  \cite{alex}, vibrational properties of fractal structures \cite{ram}, weakly
  disordered systems \cite{wl} and, more recently, quantum chaos \cite{smil}.  
  Of course, mathematicians \cite{roth} are also interested in that subject.
  
 Let us come back, for the moment, to the physics of disordered systems.
   In \cite{mont}, the authors emphasized the central role played by 
    the spectral determinant of the Laplacian in the computation of the 
  weak localization corrections. By constructing the Green's function on the
  graph, they obtained a compact form for this determinant.
  (See also 
   \cite{akk} where a path integral approach is developed; in particular,
   a trace formula for the Laplacian on a graph \cite{roth} is recovered). 

  Recently, the result of \cite{mont} was generalized \cite{jd,jd1}
  to the spectral determinant \ $\det (H + \gamma) \; (\equiv S(\gamma ))$
  with \ $H = -\Delta + V(x) $. $V(x)$ is some external 
     potential defined in each point $x$ of the graph and $\gamma$ is a
  constant (spectral parameter). In \cite{jd1}, the computation was done with
 the help of a  path integral representation of the spectral determinant 
  and also using time-dependent harmonic oscillator properties.   
        Schr\"odinger operators have also been considered 
	in \cite{texier} where the scattering matrix is computed for graphs
	made of one-dimensional wires connected to external leads.

\vskip.3cm

 All this was done assuming continuity of the eigenfunctions at each vertex.
 
\vskip.3cm

  Nevertheless, this ``natural''(!) assumption is highly questionable.
 
 For instance, in  \cite{avr,ex}, the authors argue that the reduction of
  a realistic system of coupled tubes to a graph model is far from being
  obvious. In particular, serious problems arise from the finite thickness
  of the  tubes, the geometry of the connection regions and also from
  eventually  applied external fields. Analyzing in details a model of
  junction (what they call the ``geometric-scatterer junction''), 
  they suggest that it would be more appropriate  to consider
  general boundary  conditions on the resulting graph.

\vskip.2cm

 This is the point of view we will take up in this paper when computing the
 spectral determinant\footnote{Spectral properties of graphs
 with general -- even random -- boundary conditions imposed at the vertices 
 have already been studied in the context of  quantum chaos 
   \cite{kottos}.}.
  Moreover, we will show that ``playing''
 with the  boundary  conditions allows us to study some properties of closed
 random walks on any graph. For instance, it is possible
  to count the number of such walks
  when the  number of backtrackings on each of them is fixed 
  \cite{ih}-\cite{wu}.

\vskip.5cm

 The paper is organized as follows.
In section \ref{s2}, we set up the notations that will be used throughout
the paper. Section \ref{s3} will be devoted to the computation of the
 spectral determinant $S(\gamma )$ 
  for general boundary conditions. Another expression  for  $S(\gamma )$
  will be derived in section  \ref{s4} for permutation-invariant conditions.
   Applications to countings of restricted random walks on any graph will be
   discussed at the end of this section. Finally, a short conclusion will be
   given in section  \ref{s5}.

\section{Definitions and notations}\label{s2}

  We consider a graph $ \cal G $ made of $V$ vertices,
   numbered from $1$ to $V$, linked by $B$ bonds of finite lengths.
   The coordination  
of vertex $\alpha$ is   $m_{\alpha }$ ($\sum_{\alpha =1 }^V m_{\alpha }= 2 B$).

\vskip.2cm

 On each bond [$\alpha\beta$], of length
     $l_{\alpha\beta }$, we define the coordinate $x_{\alpha\beta }$ that
     runs from $0$ (vertex $\alpha$) to $l_{\alpha\beta }$ (vertex $\beta$).
     We will also use 
     $\, x_{\beta\alpha } = l_{\alpha\beta }- x_{\alpha\beta }       $.

\vskip.2cm

  To avoid cumbersome notations, $\Phi$ being some function defined on the
     graph,  we will     simply write   $\, \int_{[\alpha\beta ]}\Phi \, $ for 
$\, \int_0^{l_{\alpha\beta }}
 \Phi (x_{\alpha\beta }) \; \d x_{\alpha\beta } \, $.  

\vskip.2cm

 An arc  ($\alpha\beta$) is defined as the oriented bond
  from $\alpha$ to $\beta$.
 Each bond  [$\alpha\beta$]
 is therefore associated  with two arcs
    ($\alpha\beta$) and  ($\beta\alpha$). In the sequel, we will consider the
  following ordering of the $2B$ arcs: 
 $ (1\alpha_1)(1\alpha_2)\ldots (1\alpha_{m_1})(2\beta_1)\ldots
 (2\beta_{m_2})\ldots  $ 

\vskip.6cm

 Concerning the eigenfunctions $\phi$ of $H$  on the graph, we define on 
  each link  [$\alpha\beta$]: 
\bea
 \fal \equiv \lim_{\xab\to 0} \phi (\xab )  \quad &;& \quad
 \fbe \equiv \lim_{\xba\to 0} \phi (\xba ) \label{d1} \\
  \fpal \equiv \lim_{\xab\to 0}\frac{\d \phi (\xab ) }{\d \xab }   \quad &;&
  \quad
  \fpbe \equiv \lim_{\xba\to 0}\frac{\d  \phi (\xba ) }{\d \xba }  \label{d2} 
\eea
($\fpal$ is the outgoing derivative at vertex $\alpha$ along the arc
  ($\alpha\beta$)).

\vskip.2cm

For the Green's function $G(x,y)$ ($x\in  [\alpha\beta ]$,
 $y$  anywhere on the graph), we similarly define: 
\bea
 \Gab (y) \equiv \lim_{\xab\to 0}  G (\xab ,y )  \quad &;& \quad
 \Gba (y) \equiv \lim_{\xba\to 0}  G (\xba ,y) \label{d3} \\
  \Gpab (y)\equiv \lim_{\xab\to 0}\frac{\d G (\xab ,y)}{\d \xab }  \quad &;&
  \quad
  \Gpba (y)\equiv \lim_{\xba\to 0}\frac{\d G (\xba ,y)}{\d \xba }  \label{d4} 
\eea 

\vskip.4cm

With the above quantities, we can build the four ($2B \times 1$) vectors
 $\phi$, $\phi'$, $G(y)$ and $G'(y)$, respectively of components
  $\fal$,   $\fpal$,   $\Gab (y)$ and  $\Gpab (y)$.

\vskip.6cm

In those conditions, the generalized boundary conditions for the operator
 $H$ on the graph can be written:
 
\be\label{bc} 
   C \; \phi \; + \; D\; \phi' \; = \; 0 
\ee
where $C$ and $D$ are two ($2B \times 2B$) constant matrices that don't
depend on $\gamma $. 

\vskip.6cm

In \cite{ks}, the authors established the conditions
 for the operator $H$ to be self-adjoint:
 $CD^+$ must be self-adjoint and the ($2B \times 4B$) matrix ($C,D$)
 must have maximal rank $2B$.

\vskip.6cm

 Local boundary conditions  connect, for each vertex
 $\alpha$, the $\phi_{(\alpha\beta_i)}$'s to the  $\phi'_{(\alpha\beta_j)}$'s,
   $i,j=1,\ldots ,m_{\alpha }$. For such conditions, $C$ and $D$ can be chosen
  block-diagonal, the square block $C_{\alpha }$ (or $D_{\alpha }$)
 being of dimension $m_{\alpha }$. If, in addition,
 we assume that, for each vertex $\alpha$, the
 conditions are invariant in any permutation of the nearest neighbours of
 $\alpha$, we can write:
 
\bea     
   C_{\alpha } &=&  c_{\alpha } {\bf 1}  + t_{\alpha }  F_{\alpha }
   \label{bci} \\ 
   D_{\alpha } &=&  d_{\alpha } {\bf 1}  + w_{\alpha }  F_{\alpha }
   \label{bci1} 
\eea

\noindent
where {\bf 1} is the unit matrix and  $F_{\alpha }$ is a matrix with all its
elements equal to $1$.  The constants
$c_{\alpha }$, $d_{\alpha }$, $t_{\alpha }$ and $w_{\alpha }$ characterize
 the boundary conditions in $\alpha$. 

 Remark that
$c_{\alpha }$ and $d_{\alpha }$ can't  both  vanish because of the maximal
rank condition\footnote{Moreover, this condition imposes that, at least,
 one of the two matrices,  $C_{\alpha }$ or  $D_{\alpha }$, is invertible
 and can be set equal to  {\bf 1} because of the homogeneity of condition
 (\ref{bc}). Finally, self-adjointness of  $CD^+$ implies that only two
 real  
parameters are, actually, necessary to characterize the boundary conditions 
   at each vertex $\alpha$ \cite{exner}. This will appear explicitly in section
    \ref{s42} where our results are expressed in terms of the two parameters
  $\eta_{\alpha}$ and   $\rho_{\alpha}$ defined in
  eqs.(\ref{p800}, \ref{p801}).}. 

\vskip.2cm

It is easy to realize that the quantity
$c_{\alpha }\phi_{(\alpha\beta_i)} +  d_{\alpha }\phi'_{(\alpha\beta_i)}   $
 doesn't depend on $i$, i.e. it is the same for all the arcs starting
 at $\alpha$. To conclude this section, let us mention the two 
  limiting cases:

\vskip.1cm

i) $d_{\alpha }=0$ that ensures the continuity of $\phi$ at vertex $\alpha$
 (thus $\phi (\alpha )$ is defined) and leads to
  $$ \sum_{j=1}^{m_{\alpha }}\phi'_{(\alpha\beta_j)} =
   -\left( \frac{ c_{\alpha } + m_{\alpha } t_{\alpha }  }
   { w_{\alpha } }\right) \phi (\alpha )
   \equiv  \lambda_{\alpha } \phi (\alpha ) $$
($\lambda_{\alpha }=0$ corresponds to  Neumann boundary conditions).  

\vskip.1cm

ii)  $c_{\alpha }=0$. In that case, all the outgoing derivatives in $\alpha$ 
 are equal.

\vskip.5cm

Now, we turn to the computation of the spectral determinant
   $S(\gamma ) ( \equiv \det (H+\gamma ))$  of the operator 
   $H = -\Delta + V(x) $ defined 
 on the graph with boundary conditions given by (\ref{bc}).

\section{General boundary conditions}\label{s3}

 As in \cite{mont,jd}, we construct  the Green's function $G(x,y)$
 on the graph:

\be\label{11}
(\gamma +H)\, G(x,y)=\delta(x-y)
\ee
and use the relationship:

\be\label{12} 
\int_{\rm Graph} G(x,x) \d x = \dg \ln \det(H+\gamma)
\ee 

\vskip.3cm

  In this section, we will consider, for each bond 
 [$\alpha\beta$], two independent solutions, $\pal$ and $\pbe$, 
  of the equation: 
    
\be\label{1}  
     (H+\gamma)\, \psi = 0 
\ee   

Those functions are chosen to satisfy:
\be\label{2}
 \pal (\alpha)=1  \quad ;\quad \pal (\beta) =0 
\ee

\be\label{3}
 \pbe (\alpha)=0  \quad ;\quad  \pbe (\beta) =1 
\ee

Their wronskian may be presented as:
\be\label{4}
\wal \equiv \pal\dpt -\pbe\dpl = \dpt (\alpha) = - \dpl (\beta) = \wla
\ee

We also define:

\be\label{n3}
\ppal (\alpha ) \equiv \frac{ \d \pal }{ \d \xab } (\xab = 0) \quad ;
 \quad \ppbe (\beta ) \equiv \frac{ \d \pbe }{ \d \xba } (\xba = 0)
\ee

\vskip.4cm

So, let us construct this Green's function  $G(x,y)$.
We first suppose that  $y$ belongs to some link $[ab]$. 

\vskip.4cm

If $x$ is located on another  bond $[\alpha\beta ]$,  we have:

\be\label{13}
G(x,y)=\Gab (y)\, \pal (x) +  \Gba(y)\, \pbe(x) 
\ee

 Taking the derivative in $\alpha$ and using (\ref{4},\ref{n3}), we get: 

\be\label{n13}
 \Gpab (y)= \Gab (y)\, \ppal (\alpha ) +  \Gba (y)\, \wal 
\ee

\vskip.4cm

 On the other hand, if $x$   belongs to the same bond $[ab]$ as $y$,
  $G(x,y)$ must satisfy, when $\epsilon\to 0$:
\bea\label{14} 
 G(y-\epsilon,y)&=& G(y+\epsilon,y)           \\
\label{15}
 \frac{ \d G}{ \d x}\bigg\vert_{x=y-\epsilon} 
 &=& \frac{ \d G}{ \d x}\bigg\vert_{x=y+\epsilon}+1 
\eea

This leads to:
\bea
x\le y \quad\quad G(x,y)= \Gaa (y)\,\pab (x)+\Gbb (y)\,\pba (x)&+&
\frac{\pab(y)\,\pba(x)}{\wab} \label{16}  \\
x\ge y \quad\quad G(x,y)= \Gaa (y)\,\pab (x)+\Gbb (y)\,\pba (x)&+&
\frac{\pba(y)\,\pab(x)}{\wab} \label{17}
\eea
($x<y \,$  means that point $\, x\, $ is closer to $\, a\, $ than $\, y\,$).

\vskip.2cm

For the derivative in $a$, we obtain:

\be\label{n16}
 \Gpaa (y)= \Gaa (y)\, \ppab (a ) +  \Gbb (y)\, \wab + \pab (y) 
\ee

Eqs. (\ref{n13},\ref{n16}) can be written in matrix form:
\be\label{n160}
 G'(y)= N G(y) - L(y)
\ee
where $N$ is a ($2B \times 2B$) square matrix with elements:
\be\label{n161}
  N_{(\alpha\beta )(\mu\eta )}  = \delta_{\alpha\mu } 
 \delta_{\beta\eta  } \ppal (\alpha ) +  \delta_{\alpha\eta } 
 \delta_{\beta\mu  } \wal
\ee
and $L(y)$ is a ($2B \times 1$) vector:
\be\label{n1610}
  L(y)_{(\alpha\beta )}  =-\left( \delta_{\alpha a } 
 \delta_{\beta b  } \pab (y ) +  \delta_{\alpha b } 
 \delta_{\beta a  } \pba (y) \right)
\ee

Eq.(\ref{n160}) together with the boundary condition (\ref{bc}) lead to:
\bea
 G(y) &=& T \; L(y) \label{n162} \\
 \mbox{with the square matrix \ } \quad 
 \quad T &=& (C+DN)^{-1}D  \label{n163}
\eea

We deduce:
\bea
 \Gaa (y) &=& T_{(ab)(ab)} L(y)_{(ab)} +  T_{(ab)(ba)} L(y)_{(ba)}
 \label{n164} \\
 \Gbb (y) &=& T_{(ba)(ab)} L(y)_{(ab)} +  T_{(ba)(ba)} L(y)_{(ba)}
 \label{n165}
\eea
and, after simple manipulations:
\bea
G(y,y) = T_{(ab)(ab)}\left( -\pab^2(y)   \right) &+& T_{(ba)(ba)}
\left( -\pba^2(y)  \right)  \, +    \nonumber \\
     &+&
\left(- T_{(ab)(ba)} - T_{(ba)(ab)} + \frac{1}{\wab} \right)
\left( \pab(y)\,\pba(y) \right)    \label{n166}
\eea

To take the trace of $G$, we must first  integrate $\pab^2(y)$,
 $\pba^2(y)$ and  $\pab (y)\pba (y)$ on $[ab]$. We have shown in 
  \cite{jd} that:
\bea\label{26} 
   \int_{[ab]} \pab^2 &=& - \dg \ppab (a)       \\
\label{27}
   \int_{[ab]} \pba^2 &=&  - \dg \ppba (b)      \\
\label{28}
   \int_{[ab]} \pab\,\pba &=& - \dg \wab
\eea

\vskip.2cm

Thus:

\bea    
\int_{[ab]} G(y,y)=T_{(ab)(ab)} \dg \ppab (a)   
 + T_{(ba)(ba)}  \dg \ppba (b) 
 &+& \left( T_{(ab)(ba)} + T_{(ba)(ab)}   \right)  \dg \wab \ -  \nonumber \\
 &-& \frac{1}{\wab }  \dg \wab \label{39}
\eea

\vskip.2cm

Now, we sum over all the bonds. With the definitions of the
 matrices $N$ and $T$,  we obtain ($C$ and $D$ don't depend on $\gamma$):

\be\label{40}
\int_{\rm Graph} G(y,y) = {\rm Tr} \left( (C+DN)^{-1}\dg (C+DN)\right) -
  \dg \left(\sum_{[ab]} \ln \wab  \right)
\ee

Finally, with the observation that 
$\, {\rm Tr}((C+DN)^{-1}\dg (C+DN)) =\dg\ln\det (C+DN)$,
 we get the spectral determinant (up to an 
 inessential multiplicative constant):

\be\label{n40} 
 S(\gamma ) \equiv \det (H+\gamma ) = 
 \prod_{[\alpha\beta ]}  \frac{1}{ \wal }\ \det(C+DN)
\ee
where  $  \prod_{[\alpha\beta ]}  \ldots   $ means product over all the bonds.

\vskip.4cm

The expression (\ref{n40}) is valid for quite general (even non-local) boundary
conditions\footnote{For local coupling, $C$
 and $D$ are block-diagonal matrices. $N$ is always  block-diagonal
  but not built with the same blocks as  $C$ and $D$. Thus, 
    no 
   further simplification appears in that case 
   when evaluating  $\det(C+DN)$ in eq.(\ref{n40}).}.

\vskip.5cm

It is worthwhile to mention that eq.(\ref{n40}) can be, heuristically,
recovered by a path integral approach 
(see \cite{akk}  (\cite{jd1}) for the Neumann case without (with) an
external potential). We will not use this way in the present work.

\vskip.5cm

Coming back to (\ref{n40}) and introducing the matrix $R$:
\be\label{n41} 
  R \equiv  ( \sg {\bf 1} + N )( \sg {\bf 1} - N  )^{-1}
\ee
 we get:
\bea
 \det (H+\gamma ) &=&  \prod_{[\alpha\beta ]} \frac{1}{ \wal } \;
 \frac{1}{\det ( {\bf 1} + R )} \; \det (C - \sg D ) \;
 \det( {\bf 1} - Q R ) \label{n42} \\
  Q  &=&   (  \sg D  - C  )^{-1}(  \sg D  + C     )   \label{n43}
\eea

\vskip.3cm

Let us remark that, for the free case ($V(x)\equiv 0$),
the matrix $R$ is quite  simple. Indeed, in that case:

\bea
  N_{(\alpha\beta )(\mu\eta )}  &=& \delta_{\alpha\mu } 
 \delta_{\beta\eta  } \coth \sg\lab +  \delta_{\alpha\eta } 
 \delta_{\beta\mu  } \left( \frac{-1}{\sinh\sg\lab} \right) \label{n44} \\
  R_{(\alpha\beta )(\mu\eta )}    &=& 
  \delta_{\alpha\eta }  \delta_{\beta\mu  }  e^{-\sg \lab }    \label{n45} 
\eea

\noindent
The matrix $R$ couples any arc ($\alpha\beta$) to its time-reversed
($\beta\alpha$).
Those considerations will show useful at the end of this paper.

\vskip.6cm

Now, let us show  that, for permutation-invariant boundary conditions
 (and $V(x)\ne 0$), the spectral determinant can be expressed in terms
  of a vertex ($V \times V$) matrix.

\section{Permutation-invariant boundary conditions}\label{s4}

The boundary conditions are now given by eqs.(\ref{bci},\ref{bci1}).

\subsection{With an external potential}\label{s41}

To compute  the spectral determinant,  we will proceed as before but,
  this time, we will consider, for each bond, two other
  independent solutions, $\cab$ and $\cba$, of the equation
    $ (H+\gamma)\, \chi = 0 $.   They  are chosen, now, to satisfy
  the following conditions:

\bea
 c_{\alpha }\cab (\alpha ) + 
 d_{\alpha }\frac{ \d \cab }{ \d \xab } (\alpha ) &=& 1        \label{p1}  \\
 c_{\beta }\cab (\beta ) + 
 d_{\beta }\frac{ \d \cab }{ \d \xba } (\beta ) &=& 0        \label{p2}  \\
 c_{\alpha }\cba (\alpha ) + 
 d_{\alpha }\frac{ \d \cba }{ \d \xab } (\alpha ) &=& 0        \label{p3}  \\
 c_{\beta }\cba (\beta ) + 
 d_{\beta }\frac{ \d \cba }{ \d \xba } (\beta ) &=& 1        \label{p4}  
\eea

As before, we  will set: 
$$ \frac{ \d \cab }{ \d \xab } (\alpha ) \equiv \cpab  (\alpha ) \quad ; \quad
   \frac{ \d \cba }{ \d \xba } (\beta )   \equiv \cpba  (\beta )  $$

The wronskian    of $\cab$ and $\cba$ writes:

\be\label{p5}
\Wal \equiv \cab \frac{ \d \cba }{ \d \xab }  - 
   \cba \frac{ \d \cab }{ \d \xab }       = \Wla
\ee

With eqs.(\ref{p1})-(\ref{p4}), we get the useful  relations:

\be\label{p6}
  c_{\alpha }\Wal =  \frac{ \d \cba }{ \d \xab } (\alpha ) \quad ; \quad 
   d_{\alpha }\Wal =   -    \cba (\alpha )
\ee

\vskip.5cm

Let us show what happens for the Green's function $G(x,y)$.
 We still assume  $y \in [ab]$.

\vskip.2cm

  For $x\in [\alpha\beta ] \ne [ab]$, we write:

\be\label{p7}
G(x,y)= B_{(\alpha\beta )}(y) \cab (x) +  B_{(\beta\alpha )}(y) \cba (x)
\ee
where  the quantities $B_{(\alpha\beta )}(y)$ are to be determined.

\vskip.2cm

Of course, if $x\in [ab]$, an additionnal term of the form
``$\caa\cbb/\Wab$'' must appear (see eqs. \\ (\ref{16} ,\ref{17})).
  
\vskip.2cm

Nevertheless, with the boundary conditions (\ref{bci},\ref{bci1}),
 it can be shown 
 that, for any vertex $\alpha $, the quantity  $B_{(\alpha\beta_i )}(y)$ 
 where $\beta_i $ is a nearest neighbour of $\alpha $,
does not  depend on $i$. In those conditions, we can set: 
 $B_{(\alpha\beta_i )}(y) \equiv B_{\alpha }(y ) $ and write for the 
 Green's function:

\bea
i) \; x\in  [\alpha\beta ] \ne [ab]  & &    \nonumber   \\
 G(x,y) &=& B_{\alpha }(y)\cab (x) + B_{\beta }(y)\cba (x) \label{p8} \\
ii) \; x\in [ab] \qquad \quad  & &      \nonumber  \\
   x\le y \quad        G(x,y) &=& B_a(y)\caa (x) + B_b(y)\cbb (x)
 + \frac{\caa (y) \cbb (x)}{\Wab }  \label{p9} \\
   x\ge y \quad        G(x,y) &=& B_a(y)\caa (x) + B_b(y)\cbb (x)
 + \frac{\cbb (y) \caa (x)}{\Wab }  \label{p10} 
\eea

\vskip.3cm

The boundary conditions lead to the equation:

\be\label{p11}
  M \; B \; = \; {\cal L}
\ee
where $M$ is a ($V \times V$) matrix with elements:
\bea 
 M_{\alpha\alpha } &=& 1+ t_{\alpha }\left(
 \sum_{i=1}^{m_{\alpha }} \cabi (\alpha )   \right) +
  w_{\alpha }\left(
 \sum_{i=1}^{m_{\alpha }} \cpabi (\alpha )   \right) \label{p12}   \\
 M_{\alpha\beta } &=& ( c_{\alpha } w_{\alpha } -  t_{\alpha } d_{\alpha } )
 \Wal   \qquad  \mbox{ \ if $[\alpha\beta ]$ is a bond}   \label{p13}    \\
                &=& 0 \qquad \ \mbox{ otherwise} \nonumber
\eea

$B$  and ${\cal L}$ are two ($V \times 1$) vectors of components:

\bea
         B_{\alpha } &=&  B_{\alpha }(y) \label{p14}  \\
      {\cal L}_{\alpha } &=&  -\left(   \delta_{\alpha a } 
  \caa (y ) (c_a w_a - d_a t_a ) +  \delta_{\alpha b } 
  \cbb (y) (c_b w_b - d_b t_b )  \right) \label{p15}
\eea

Solving (\ref{p11}) and taking the trace of $G$  with the relations \cite{jd}:
\bea 
  d_a  \int_{[ab]} \caa^2 &=&  \dg \caa (a)  \label{p16}     \\
  d_b  \int_{[ab]} \cbb^2 &=&  \dg \cbb (b)  \label{p17}     \\
  \int_{[ab]} \caa\, \cbb &=& - \dg \Wba     \label{p18}
\eea
 we finally get the spectral determinant
 (still up to a multiplicative constant):

\be\label{p19} 
 \det (H+\gamma ) =  \prod_{[\alpha\beta]}\frac{1}{ \Wal }\ \det(M)
\ee

Comparing the asymptotic behaviours of the right-hand sides of
 (\ref{n40}) and (\ref{p19})
 when $\gamma\to\infty$, we establish the following equality that is valid in
 the presence of a potential $V(x)$ and 
  for permutation-invariant boundary conditions: 
\be\label{p20} 
 \prod_{[\alpha\beta ]}\frac{1}{ \wal }\ \det(C+DN)
 =  \prod_{[\alpha\beta ]}\frac{1}{ \Wal }\ \det(M)
\ee
Recall that, for such boundary conditions, $C$ and $D$ are block-diagonal 
 matrices given by eqs.(\ref{bci}, \ref{bci1}).

\subsection{Free case}\label{s42}

Let us study the case $V(x) \equiv 0$ still with 
 permutation-invariant boundary conditions.

\noindent
With the notations 
\bea
    \eta_{\alpha } &=& \frac{c_{\alpha }   +  \sg d_{\alpha } }
                            { c_{\alpha }  -  \sg d_{\alpha } } \label{p800} \\
    \rho_{\alpha } &=& \frac{\mu_{\alpha }^- - \mu_{\alpha }^+ }
                            {1+ m_{\alpha } \mu_{\alpha }^- } \label{p801} \\
    \mu_{\alpha }^{\pm } &=& \frac{t_{\alpha }   \pm  \sg w_{\alpha } }
      { c_{\alpha }  \pm  \sg d_{\alpha } } \label{p802} 
\eea
(\ref{p20}, \ref{n42}, \ref{n43}) lead to:

\be\label{p804} 
\det ({\bf 1} - Q R) = 2^{-V} \prod_{\alpha } \left( \rho_{\alpha }
\eta_{\alpha } \right) \prod_{[\alpha\beta ]}\left( 
 1- \eta_{\alpha } \eta_{\beta } e^{- 2 \sg \lab}   \right)  \; 
 \det \widetilde{M} 
\ee
with the ($V \times V$) $ \; \widetilde{M} $ matrix:

\bea
       \widetilde{M}_{\alpha\alpha }        &=&
 \frac{2}{ \rho_{\alpha }\eta_{\alpha } } - 
 \frac{ m_{\alpha }}{\eta_{\alpha } } +  \frac{1}{ \eta_{\alpha }} 
\sum_{i=1}^{ m_{\alpha }}\left( 
\frac{1+\eta_{\alpha }\eta_{\beta_i }e^{-2\sg l_{\alpha\beta_i}} }
{  1-\eta_{\alpha }\eta_{\beta_i }e^{-2\sg l_{\alpha\beta_i}}      }
 \right)  \label{p805}  \\
    \widetilde{M}_{\alpha\beta }        &=&
   \frac{-2  e^{- \sg l_{\alpha\beta }   }}
   {   1-\eta_{\alpha }\eta_{\beta }e^{-2\sg l_{\alpha\beta }}       }
   \qquad \mbox{if [$\alpha\beta $ ] is a bond}     \label{p8050}  \\
         &=&    0   \qquad \qquad   \mbox{otherwise} \nonumber
\eea

For permutation-invariant boundary conditions, the matrices $C$, $D$ and
 $Q$  (eq.(\ref{n43})) are block-diagonal. The block $Q_{\alpha }$ takes the
 simple form:
\be\label{p900}
     Q_{\alpha } = \eta_{\alpha } 
     \left( -{\bf 1} + \rho_{\alpha }  F_{\alpha } \right)
\ee

The only non-vanishing elements of the $QR$ matrix are:

\be\label{p901}
    (QR)_{(\alpha\beta ) (\mu\alpha )}= \left(
    \rho_{\alpha }\eta_{\alpha } - \eta_{\alpha } \; \delta_{\beta\mu } 
 \right) \; e^{- \sg l_{\alpha\mu }} 
\ee

\vskip.3cm

 In view of the following application, we will say that
 $ \rho_{\alpha }\eta_{\alpha } - \eta_{\alpha }  $
 is the reflection factor in $\alpha $ and 
 $ \rho_{\alpha }\eta_{\alpha }   $ is the transmission factor.

\vskip.5cm

Expanding

\be\label{p8000}
\ln \det ({\bf 1} - QR )=  - \sum_{n=1}^\infty \frac{1}{n} \; 
 \mbox{Tr} \; (QR)^n  
\ee
and following the development of  \cite{akk}, we finally get:
 
\vskip.1cm 

\be\label{p902}
 \det( {\bf 1} - QR ) = \prod_{\widetilde{C}} \left( 1- \mu(\widetilde{C} )
 e^{-\sg l(\widetilde{C} )} \right)
\ee
where the product is taken over all primitive orbits $\widetilde{C}$.
 Recall that an orbit is said to be primitive if it cannot be decomposed as a
  repetition of any smaller orbit. $l(\widetilde{C} )$ is the length of
 $ \widetilde{C} $.  

 An orbit being a succession of arcs \ 
 $\ldots  (\tau\alpha )(\alpha\beta  )  \ldots $ \ with, in $\alpha $,
  a reflection (if $\tau =\beta $) or a transmission  (if $\tau \ne \beta $),
  the weight $\mu(\widetilde{C} )$, in eq. (\ref{p902}), will be the product of
  all the reflection -- or transmission -- factors along $\widetilde{C}$.

\vskip.5cm

 Henceforth, we will consider the situation where the spectral
 parameter $\gamma$ is equal to $1$ and, in addition:
$$  \rho_{\alpha }\eta_{\alpha }=1 \quad ; \quad  \eta_{\alpha }=\eta \quad
 ; \quad  \lab = l  $$ 
for all the vertices and bonds of the graph. 

\vskip.5cm

With  $u \equiv e^{-l}$, (\ref{p804}) takes the simple form:

\be\label{p903}
\prod_{\widetilde{C_m}} \left( 1 - (1-\eta )^{n_R(\widetilde{C_m} )} u^m
 \right) = (1- \eta^2 u^2 )^{B-V} \det 
 \left( ( 1- \eta^2 u^2  ) {\bf 1} +\eta u^2 {\bf Y} - u {\bf A}   \right)
      \left(  \equiv  Z^{-1}  \right)
\ee

\noindent
$m$ is the number of steps of the primitive orbit $\widetilde{C_m}$ and
 $n_R(\widetilde{C_m})$ is the number of reflections (backtrackings) occuring
  along  $\widetilde{C_m}$. 

   {\bf Y} is a ($V\times V$) matrix with elements \ 
 $Y_{\alpha\beta }=\delta_{\alpha\beta } \; m_{\alpha }$ \ and {\bf A} is the
  adjacency matrix ($A_{\alpha\beta }=1$ if [$\alpha\beta $]
 is a bond, $=0$ otherwise).
 
\vskip.2cm

Setting $\eta=1$ implies $n_R(\widetilde{C_m})=0$ in the left-hand side of
 (\ref{p903}): we recover Ihara's formula \cite{ih,terras} where only 
  primitive orbits without tails and backtrackings are kept. (Ihara
 \cite{ih}  established this formula for a regular graph;
 the proof for a general
  graph is done in \cite{terras}  using a direct - and somewhat tedious -
  counting technique). 
 
\vskip.5cm

Now, let us consider random walks with a given number of backtrackings.
 Eqs.(\ref{p8000}) and (\ref{p901}) suggest an expansion in closed random
 walks on the graph. Taking $Z$ in (\ref{p903}), we get: 

\be\label{p1949}
u \; \frac{ \d \ln Z}{ \d u} = \sum_{m=2}^{\infty } \; \sum_{p=0}^m \; 
\sum_{\alpha =1}^V \; N_m^p(\alpha ) \; (1-\eta )^p \; u^m 
\ee
where $ N_m^p(\alpha ) $ is the number of $m$-steps closed 
random walks on the graph starting at $\alpha$, with $p$ backtrackings.

\vskip.4cm

For the complete graph (each vertex $\alpha$ is linked to all the other
 vertices of the graph), we get the results:
 
\bea
   N_2^0(\alpha )      &=& 0         \nonumber \\
  N_3^0(\alpha )      &=&   (V-1)(V-2)      \nonumber \\
   N_4^0(\alpha ) &=&    (V-1)(V-2)(V-3)       \nonumber \\
   N_5^0(\alpha ) &=&      (V-1)(V-2)(V-3)(V-4)     \nonumber \\
       N_6^0(\alpha ) &=&   (V-1)(V-2)(V^3-9V^2+29V-32)  \label{p1950} 
\eea
and also:
\bea
 N_2^1(\alpha ) &=&  N_3^1(\alpha ) \ = \ N_4^1(\alpha ) \ = \ 0   \nonumber \\
   N_5^1(\alpha ) &=&      5(V-1)(V-2)(V-3)     \nonumber \\
       N_6^1(\alpha ) &=&   6(V-1)(V-2)(V-3)^2  \label{p1952} 
\eea

In \cite{wu}, the same problem is studied with probabilistic methods but for
open random walks. Closed walks are therefore obtained by identifying 
 the starting and ending points but nothing is said about an eventual
 backtracking occuring at that point. So, the results of \cite{wu} (let us call
  them  $ {\cal N}_m^p(\alpha ) $) will, in general, differ from ours.
 For instance, we checked for the complete graph, the relationship:
 
\be\label{p1951} 
   {\cal N}_m^0(\alpha ) =   N_m^0(\alpha ) + \frac{1}{m}  N_m^1(\alpha )
\ee 
(This comes from the complete symmetry of this graph).

\section{Conclusion}\label{s5}

 We  have computed the spectral determinant for a Schr\"odinger operator on
 a graph with quite general boundary conditions. The result is expressed in
 terms of an arc matrix. When the conditions are permutation-invariant,
  another expression can be derived in terms of a vertex matrix. Comparison
  of both expressions allowed us to study reflection properties of random walks
  on any graph.

\vskip.4cm

 The expansion (\ref{p902})  
 of the spectral determinant in periodic orbits is the basis
 for obtaining a trace formula (see, for instance, \cite{akk}
 where this is done in great details for Neumann boundary conditions).
 Unfortunately, in the general case, the reflection and transmission factors
 are $\gamma$-dependent and technical difficulties prevent from getting a
 trace formula in an appealing form. So, this problem is still  an open one.

\vskip1.5cm

I acknowledge Pr A Comtet and Dr C Texier for stimulating discussions.


\vskip1.5cm




\begin{thebibliography}{10}

\bibitem{ru}

K. Rudenberg and C. Scherr,  
\newblock  J. Chem. Phys. {\bf 21}, 1565 (1953).



\bibitem{alex}

S. Alexander,  
\newblock Phys. Rev. B {\bf 27}, 1541 (1983).

\bibitem{ram}

R. Rammal,
\newblock J. Phys. I (France) {\bf 45}, 191 (1984).


\bibitem{wl}

B. Dou{\c c}ot   and R. Rammal,
\newblock Phys. Rev. Lett. {\bf 55}, 1148  (1985);  
\newblock J. Physique {\bf 47}, 973 (1986); 
G. Montambaux, 
\newblock  in  {\it Proceedings of the Les Houches Summer School,
  Session LXIII},   edited by S. Reynaud, E. Giacobino,
  and J. Zinn-Justin  (Elsevier, Amsterdam, 1996)   p. 387.
 


\bibitem{smil}

T. Kottos  and U. Smilansky, 
\newblock Phys. Rev. Lett. {\bf 79}, 4794 (1997);  
\newblock Ann. Phys. (N.Y.) {\bf 274}, 76 (1999). 

  
\bibitem{roth}

J.P. Roth, 
\newblock C.R. Acad. Sc. Paris {\bf 296}, 793 (1983).




\bibitem{mont}

M. Pascaud  and G. Montambaux, 
\newblock Phys. Rev. Lett. {\bf 82}, 4512 (1999);  
M. Pascaud, 
\newblock Ph.D. thesis, Universit\'e Paris XI, 1998. 


\bibitem{akk}

E. Akkermans, A. Comtet, J. Desbois, G.  Montambaux and C. Texier, 
\newblock Ann. of Phys. {\bf 284}, 10 (2000). 



\bibitem{jd}

J. Desbois,
\newblock J. Phys. A {\bf 33}, L63 (2000).


\bibitem{jd1}

J. Desbois, 
\newblock Eur. Phys. J.  B {\bf 15}, 201 (2000).

\bibitem{texier}

C. Texier and G. Montambaux,
\newblock cond-math/0107104.


\bibitem{avr}

J.E. Avron, P. Exner and Y. Last, 
\newblock Phys. Rev. Lett. {\bf 72}, 896 (1994).



\bibitem{ex}

P. Exner,
\newblock Phys. Rev. Lett. {\bf 74}, 3503 (1995);
\newblock  J. Phys. A {\bf 29}, 87 (1996).


\bibitem{kottos}

T. Kottos and H. Schanz,
\newblock Physica E {\bf 9}, 523 (2001).




\bibitem{ih}

Y. Ihara,
\newblock J. Math. Soc. Japan {\bf 18}, 219 (1966). 



\bibitem{terras}

H.M. Stark and  A.A. Terras, 
\newblock Adv. in Math. {\bf 121}, 124 (1996). 



\bibitem{wu}

F.Y. Wu and  H. Kunz, 
\newblock cond-mat/9812203. 



\bibitem{ks}

V. Kostrykin and  R. Schrader, 
\newblock J. Phys. A {\bf 32}, 595 (1999). 

\bibitem{exner}

P. Exner and P. \v Seba,
\newblock Rep. Math. Phys. {\bf 28}, 7 (1989).


\end{thebibliography}
\end{document}